# Observing collisions beyond the secular approximation limit

Junyang Ma[1,2], H. Zhang[1], B. Lavorel[1], F. Billard[1], E. Hertz[1], J. Wu[2*], C. Boulet[3], J.-M. Hartmann[4*] and O. Faucher[1*]

[1]Laboratoire Interdisciplinaire CARNOT de Bourgogne,
UMR 6303 CNRS-Université de Bourgogne Franche-Comté, BP 47870, 21078 Dijon, France.
[2]State Key Laboratory of Precision Spectroscopy, East China Normal University, Shanghai 200062, China.
[3]Institut des Sciences Moléculaires d'Orsay, CNRS, Université Paris-Sud, Université Paris-Saclay, Orsay F-91405, France.
[4]Laboratoire de Météorologie Dynamique/IPSL, CNRS, Ecole polytechnique, Institut polytechnique de Paris, Sorbonne Université, Ecole Normale Supérieure, PSL Research University, F-91120 Palaiseau, France.
*e-mails: olivier.faucher@u-bourgogne.fr, jmhartmann@lmd.polytechnique.fr, jwu@phy.ecnu.edu.cn

**Abstract:** Probing the limits of the secular approximation and understanding how nonsecularity affects decoherence in open quantum systems are fundamental problems of interest for various applications and natural phenomena. Here we experimentally and theoretically unveil the nonsecular dynamics in the rotational relaxation of molecules due to thermal collisions by using the laser-kicked molecular rotor as a model system. Specifically, rotational coherences in $N_2O$ gas (diluted in He) are created by two successive nonresonant short and intense laser pulses. They are systematically probed by studying the change of amplitude of the rotational alignment echo with the gas density. By interrogating the system at the early stage of its collisional relaxation, we observe a significant variation of the dissipative influence of collisions with the time of appearance of the echo, featuring a decoherence and dissipation process which is well reproduced by the nonsecular quantum master equation for modeling molecular collisions.

Energy transfer through quantum coherences plays an essential role in diverse natural phenomena and technological applications, such as human vision[1], light-harvesting complexes[2], quantum heat engines[3], and quantum information and computing[4]. The understanding of the long-lived coherence involved in these phenomena requires a detailed modeling of the system-bath interactions beyond the so-called secular and/or Markovian approximations[5,6]. Despite continuous theoretical progress on understanding nonsecular dynamics in the last decades, convincing experimental observations are still lacking, in particular when they manifest through collision-induced processes. More generally, understanding how collisions between molecules change their translational and rotational motions is a fundamental problem of physics that has numerous practical implications. Indeed, most real gas media being at non-negligible pressure, intermolecular interactions affect many observable quantities. At thermodynamic equilibrium, they play a key role in the shape of light-absorption and -scattering spectra[9] and in the transports of mass and energy[10,11], for instance. When the system has been brought out of equilibrium by some external action (for example one or several resonant or nonresonant electromagnetic excitations or a plasma discharge), collisional processes are often the main channel inducing decoherence and transfers among populations bringing it back to equilibrium[10,12]. Collisional effects have been extensively studied in the spectral domain by looking at the shape of various absorption features[9]. In the time domain, it has been recently shown that the decay of alignment echoes induced by two laser kicks[7] permits time-resolved measurements of collision-induced rotational changes at short time scales[8,13]. So far, all quantum models used for the calculation of the effects of decoherence processes due to intermolecular forces are based upon the Markovian and secular approximations[10,14-16]. These allow to obtain the density matrix $\rho(t)$ describing the evolution of the system when coherences (i.e. nondiagonal elements of $\rho$) have been created (for example by laser pulse(s), as in the present work). Recall that the Markov approximation assumes[14,15] that all collisions are complete and un-correlated (no memory effects in the system-bath history) while the secular approximation assumes[16] that the delay $\Delta t$ at which the system is observed after the creation of the coherences is much greater than the periods of oscillation of the coherences.

Here we provide the experimental and theoretical evidence that rotational alignment echoes enable to probe an open molecular system at the early stage of its collisional relaxation where the secular approximation breaks down. These echoes[7] occur after molecules have been suddenly aligned by two successive nonresonant laser excitations. A first short pulse, $P_1$, impulsively aligns some of the molecules along its polarization direction thanks to the anisotropy of the molecular polarizability[17]. This results in a peak in the alignment factor $<\cos^2(\theta)>(t)$ ($\theta$ being the angle between the molecule axis and the laser polarization direction and $<>$ denoting the quantum expectation value) that quickly vanishes due to rotational dephasing. After a delay $\tau_{12}$, a second pulse, $P_2$, is applied that induces a rephasing process leading to the creation of an echo in the alignment factor at time $\Delta t = 2\tau_{12}$ after $P_1$. The decay, with pressure, of the echo is studied in $N_2O$-He gas mixtures for various delays $\tau_{12}$. The experiments show that the dissipation of the echo is all the more rapid when the delay is long, a result in contrast with the secular model predicting delay-independent results. On the other hand, very good agreement between measurements and calculations are obtained when a nonsecular approach is used.

## Results
### Experiments
The experiment is implemented with the setup depicted in Fig. 1 and further described in the Methods section. After the molecules have been aligned by two laser kicks, the anisotropy of the medium resulting from the reorientation of the molecules is interrogated by a probe pulse. A detection is performed so that the measured signal is directly proportional to the convolution of ($<\cos^2(\theta)>(t)$ -1/3) with the temporal envelope of the probe, where 1/3 is the alignment factor obtained when molecules are randomly oriented. Figure 2a exemplifies the echo features observed in the alignment signal of linear molecules at low pressure. Like any other echo phenomenon, the dominant echo response appears at the symmetric position of $P_1$ with respect to $P_2$ along the time axis. It is followed by higher-order and imaginary echoes[18,19] and revivals. All these structures exhibit an asymmetric temporal shape consisting of a peak resulting from the alignment of the molecules along the polarization of the pump field, and a dip associated with a delocalization of the molecular axes within a plane perpendicular to the field vector.

It is only very recently that the collisional dissipation of echoes has been studied[8,13], by analyzing the evolution of their amplitudes for various delays at fixed gas density. The interest of this approach to probe the dissipation of the alignment under high-pressure conditions where standard alignment revivals cannot be used was demonstrated in Ref. 8. Indeed, revivals restrict the probing of the system to specific times that are tied to the moment of inertia of the molecule. On the contrary, the rotational echo provides much more flexibility, since its time of appearance can be tuned at will by adjusting the delay between $P_1$ and $P_2$. The echo is therefore particularly well suited for interrogating the dynamics of the system at early times, much before the first revival appears. As shown in Fig. 2b, we here look at the echoes from a point of view different from that used in Refs. 8,13, and study the reduction of their amplitudes "$S$" with increasing pressure for fixed values of $\tau_{12}$. Note that all results presented from now on have been recorded in high-pressure (up to 25 bar) mixtures of $N_2O$ gas diluted in 96% of helium. The high concentration of helium, which has been chosen for its weak nonlinear optical properties, allows overcoming the nonlinear propagation effects that would take place in high-density pure molecular gases of the same nature. A first indication of the enriched information brought by the echo with respect to the revivals is revealed by Fig. 2c, which shows measured amplitudes of five echoes and two revivals as a function of the density $d$ multiplied by the time of appearance of the considered alignment structure (i.e. the echo or the $n^{th}$ revival). Obviously, the efficiency with which collisions reduce the alignment amplitude is the same for the half and full revival (at 20.2 and 40.4 ps), while it varies significantly for the echoes observed between $\tau_{12}$ =1.61 and 8.58 ps. By repeating the measurements of Fig. 2c for several delays and gas densities, we have been able to extract (see the Methods section) density-normalized characteristic time constants of the echo amplitude decay from exponential fits (see Fig. 2c). The values reported in Fig. 3 show that the echo decay is slow at short times and becomes faster as the delay increases, before a plateau is reached around $2\tau_{12}$=10 ps. In the plateau region, the echo and the revival share about the same density-normalized characteristic decay time. Note, for comparison, that the mean time interval between successive $N_2O$-He collisions, obtained for the intermolecular distance of 3.5 Angstrom where the $N_2O$-He potential is significant (see Supplementary Note 1), is of 70 ps.amagat.

**Model**

In order to explain the difference between the short- and long-time dynamics, we performed numerical simulations of the alignment signal in the presence of collisional dissipation. Starting from equilibrium before the pulses, where the density operator $\rho(t < 0)$ is Boltzmannian and diagonal with respect to the rotational quantum numbers $J$ (principal) and $M$ (magnetic), the evolution for $t \geq 0$ is obtained from the Liouville-von Neumann equation[10,14-16]

$$\frac{d\boldsymbol{\rho}}{dt}(t) = -\frac{i}{\hbar}[\mathbf{H}_0 + \mathbf{H}_L(t), \boldsymbol{\rho}(t)] + \left(\frac{d\boldsymbol{\rho}(t)}{dt}\right)_{Coll}, \quad (1)$$

where $\hbar$ is the Planck constant divided by $2\pi$, $\mathbf{H}_0$ is the free rotational Hamiltonian, and

$$\mathbf{H}_L(t) = -\frac{1}{4}E^2(t)\Delta\alpha\cos^2\theta \quad (2)$$

describes the nonresonant interaction of the molecule with the linearly polarized field (up to an irrelevant term that does not multiply the direction cosine operator $\cos^2\theta$), with $\Delta\alpha = \alpha_{//} - \alpha_{\perp}$ the anisotropic polarizability and $E(t)$ the temporal envelop of the laser pulse. Note that, in principle, the contribution of the permanent dipole moment of $N_2O$ should be included in equation (2). However, since the excitation field used in the experiment (at 800 nm) oscillates with a period (2.7 fs) much smaller than that of the rotation of the molecule for the thermally populated levels, the associated contribution to equation (2) averages out. Assuming a Markovian model of collisions (valid here because, as discussed in the Supplementary Note 1, $N_2O$ and He interact very shortly and in a very limited range of intermolecular distance[20]), and neglecting all radiative decays since they occur on a much longer time scale, the matrix elements of the dissipation term read

$$\left(\frac{d\rho_{ij}(t)}{dt}\right)_{Coll} = -d\sum_{i',j'}\Lambda_{ij,i'j'}\rho_{i'j'}(t), \quad (3)$$

where $i$ and $j$ denote the rotational states $|J,M\rangle$ of the system and $\Lambda_{ij,i'j'}$ are the density-normalized relaxation matrix elements.

**Break down of the secular approximation**

In a first step, we use the so-called Bloch model in which dissipative effects are treated within the Markovian and secular approximations (further discussed in the Supplementary Notes 1 and 2, respectively). This corresponds to neglecting all elements of the relaxation matrix responsible for transfers among coherences (nondiagonal terms of $\boldsymbol{\rho}$) and between populations (diagonal elements of $\boldsymbol{\rho}$) and coherences. Only the $\Lambda_{ij,ij}$ and $\Lambda_{ii,i'i'}$ terms are then kept which are directly related to the rates of population transfer from $|J,M\rangle$ to $|J',M'\rangle$ (see Ref. 21 and Supplementary Note 3), including inelastic ($J$-changing) and elastic reorienting ($J$-conserving, $M$-changing) processes, as well as the pure dephasing of the coherence which describes the effect of elastic collisions that interrupt the phase of the molecule without quenching it. Under these approximations, the equation driving the decay of the coherences (i.e. for $i \neq j$) becomes

$$\left(\frac{d\rho_{ij}(t)}{dt}\right)_{Coll}^{Secular} = -d\Lambda_{ij,ij}\rho_{ij}(t). \quad (4)$$

The diagonal elements $\Lambda_{ij,ij}$ have been estimated for $N_2O$-He at room temperature (see the Supplementary Note 3.1) using the Infinite Order Sudden Approximation (IOSA). The calculations show that all these terms are positive, practically independent of $i$ and $j$ and equal to 0.013 ps$^{-1}$.amagat$^{-1}$. This means that the echo decay within the secular approximation is independent of the delay. As shown by the simulations represented by the black dashed line in Fig. 3, the secular theory well reproduces the observed decays of the echoes in the plateau region beyond about $2\tau_{12}$ =10 ps with a density normalized time constant of 1/0.013=76 ps.amagat. In contrast, the secular model does not capture the weaker efficiency of collisions at shorter times and the fact that the decrease of the echo is here all the more slow when the delay is short. This observation seriously questions the validity of the secular approximation during the early dynamics of the system.

For a simple (but realistic) picture of the limits of the secular approximation, let us assume that coherences have been instantaneously generated by a laser pulse at $t$=0. This creates nonzero nondiagonal elements $\rho_{JM,J'M}(t=0^+)$ between states $|J,M\rangle$ and $|J',M\rangle$, which, in the absence of collisions, then evolve with time according to $\rho_{JM,J'M}(t)=\rho_{JM,J'M}(t=0^+)\exp[-i(E_J-E_{J'})t/\hbar]$, with $E_J$ and $E_{J'}$ the corresponding rotational energies. Let us now assume that collisions induce transfers between coherences and, in particular, from $\rho_{J_1M_1,J'_1M_1}(t)$ to $\rho_{J_0M_0,J'_0M_0}(t)$ with a real valued rate constant. Under some approximations presented in the Supplementary Note 2, it can be shown that the relative change of $\rho_{J_0M_0,J'_0M_0}(t)$ due to transfers from $\rho_{J_1M_1,J'_1M_1}(t)$ at time $t$ after the laser pulse is

$$\text{Re}\left[\frac{\Delta\rho_{J_0M_0,J'_0M_0}(t)}{\rho_{J_0M_0,J'_0M_0}(t)}\right]=B\,\text{sinc}\left[(\omega_{J_1J'_1}-\omega_{J_0J'_0})t\right], \quad (5)$$

where Re[…] denotes the real part, $B$ is a constant, sinc[…] designates a cardinal sine, and $\omega_{JJ'}\equiv(E_J-E_{J'})/\hbar$. It is obvious from equation (5) that, as $|\omega_{J_1J'_1}-\omega_{J_0J'_0}|t$ gets larger than a few times $\pi$, the transfers between coherences become negligible when compared to the total loss. Note that the imaginary part of $\Delta\rho/\rho$ exhibits the same behavior. In order to be more quantitative, let us consider the case of $N_2O$. Since only $\rho_{JM,J'M}(t)$ with $J'=J\pm2$ contribute to the echo amplitude (see the Methods section), we below focus on such elements and limit ourselves to the most populated state $J_0$ =15 at $T$=300 K. The quantity in equation (5) is plotted in Fig. 4a for $J_0$=15, $J'_0$=17, $J'_1=J_1+2$ and $J_1$=17, 19, 21, and 23 (we keep $|J_0-J_1|$ even because of the collisional selection rule, see the Supplementary Note 3.2), where all values have been normalized to unity at $t$=0. As can be seen, transfers between coherences indeed decrease rapidly. They are significant before about 5 ps and practically negligible above 10 ps, a delay beyond which the secular approximation thus becomes valid, a behavior qualitatively consistent with the difference observed at short times between the experimental results and the secular predictions in Fig. 3.

**Successful nonsecular predictions**
In order to go further, computations of the echo decays have been carried by solving equations (1)-(3) now keeping all collisional transfers channels and thus accounting for secular and nonsecular terms. Equation (3) thus becomes

$$\left(\frac{d\rho_{ij}(t)}{dt}\right)^{\text{Nonsecular}}_{\text{Coll}}=\left(\frac{d\rho_{ij}(t)}{dt}\right)^{\text{Secular}}_{\text{Coll}}-d\sum_{i',j'\neq i,j}\Lambda_{ij,i'j'}\rho_{i'j'}(t), \quad (6)$$

where the first term on the right-hand side is given by equation (4). The required relaxation matrix elements for equation (6) have been constructed using the IOSA, as explained in the Supplementary Note 3 where the consistency with the rates used previously within the secular approximation[21-23] is demonstrated. Unfortunately, due to computer-time and -memory limitations, nonsecular calculations at room temperature, which require to include $N_2O$ $J$ values up to $J_{\text{Max}}$=60, were not tractable (which is not the case when the secular approximation is used). Hence, we carried calculations for $T$=100 K and $J_{Max}$=38, conditions for which predictions for 6 densities and 21 delays $\tau_{12}$ could be completed in a "reasonable" time. The results of this exercise, in which we used rates computed at 295 K but initial populations defined at 100 K, are displayed in Fig. 3 with a red dashed line. Note that comparing the secular results at 100 K and 300 K shows that limiting computations to $T$=100 K and $J_{Max}$=38 only slightly affects the prediction (by 10 %). Figure 3 confirms that the nonsecular terms [within the sum on the right-hand side of equation (6)], which describe the transfers among coherences as well as between populations and coherences, globally contribute to reduce the pressure-induced echo decay at short times. Analysis shows that this results from the fact that these contributions here systematically reduce the collision-induced decay of the amplitudes of the $\rho_{JM,J\pm 2M}(t)$ coherences, as exemplified in Figs. 4b and c, and thus also of the echo. In particular, for $N_2O$ diluted in He, the IOS model predicts that, immediately after the first pulse, nonsecular couplings between coherences reduce the collisional damping of these amplitudes by a factor of about two with respect to the secular behavior. Neglecting the nonsecular terms thus leads to overestimate the decay of the system. Note that a similar pattern was observed in Ref. 24 in the modeling of decoherence and dissipation in non adiabatic cis-trans photoisomerization. As time goes by, the averaging out of these terms, due to progressive dephasing of the coherences (see Fig. 4a), results in a decrease of the predicted decay time constant toward the secular plateau, in perfect agreement with the measurements. Note that the convergence of the nonsecular results toward the delay-independent secular ones (Fig. 3) explains the experimentally observed close similarity of the decays of the revivals and of the echo for large delays (Fig. 2c).

## Discussion

Collisional dynamics of rotationally excited $N_2O$ molecules diluted in helium gas has been investigated using alignment echoes. The latter enable to probe the system very soon after its stimulation by two laser kicks, on a time scale of a few ps that is comparable to the time interval between successive collisions (about 3.5 ps for the density of 20 amagat typical of the experiments) at the high pressures used in the measurements. This unprecedented scrutiny enables the observation of collisional transfers occurring in the nonsecular regime and leading to a slowing-down of the decoherence of the system lasting for a few picoseconds. This observed longevity of coherences challenges the traditional wisdom that interactions with the environment universally lead to decoherence only, and boosts the theoretical revisiting of the quantum master equations normally used to describe the system-bath interactions. It also opens renewed perspectives, through pure time-domain experiments, for our understanding of collisional processes and tests of rotational relaxation models. In particular, probing echoes at short times for collisional pairs involving significantly smaller relative speeds and longer-range intermolecular forces that does the $N_2O$-He pair studied in this work should enable to evidence the breakdown of the Markov approximation. Finally, note that equivalent studies could be made for other types of molecules (including polar and nonlinear ones) and/or carried out by using a different process in order to excite the system at $t$=0 and $t$=$\tau_{12}$ and generate an echo at $2\tau_{12}$. For instance, early time decoherence could be investigated for $CH_3I$ using THz excitations as done in Refs. [25-27] with a single pulse, which is only one example among many others of the great potentialities of the approach introduced in this paper to study early time dissipation.

## Methods
### Experimental set-up
The experiment is based on a chirped femtosecond Ti:Sapphire amplifier delivering 800 nm light pulses with a repetition rate of 1 kHz and a duration of 100 fs (FWHM). The output of this laser is first separated into a pump and a probe beam. The pump beam, used for impulsive alignment, is directed through a Mach-Zehnder interferometer producing two linearly-polarized collinearly-propagating pulses $P_1$ and $P_2$ with a time delay controlled by a motorized stage. In order to reduce the perturbation of the detection by the light of the pump pulses scattered by the molecules and the cell windows, the probe beam is frequency doubled with a type I BBO crystal before being sent through a second motorized delay line. For the purpose of the birefringence detection, its polarization is rotated by 45° with respect to the pump pulses. The pump and probe beams are both focused in a high-pressure static cell by the same lens (f=100 mm) in a noncolinear geometry leading to a crossing angle of 4° at focus. The intensities of $P_1$ and $P_2$ are estimated around 20 and 13 TW.cm$^{-2}$, respectively. After filtering out the 800 nm photons and collecting the 400 nm light with a second lens, the transmitted probe light is analyzed by a low noise balanced detection. The latter is achieved by combining a quarter-wave plate and a Wollaston prism separating the vertical and horizontal polarization components that are then independently measured by two connected head-to-tail photodiodes. All measurements were made at room temperature (295 K) for 4%$N_2O$+96%He mixtures at various total pressures up to 25 bar.

**Computations**

The density matrix $\rho$ was initialized, before the first laser pulse, to its equilibrium value at 100 K and 295 K (see text). Its evolution with time was then obtained by solving equations (1) and (2) for a 100 fs Gaussian pulse envelop (FWHM) and the same pulses intensities as in the experiments, using the anisotropic polarizability[28] $\Delta\alpha$=19.8 $a_0^3$ and the rotational constant $B = 0.419$ cm$^{-1}$ of $N_2O$. The collisional term was computed using equation (3), with rates $\Lambda_{J_1 M_1, J_1' M_1 \rightarrow J_0 M_0, J_0' M_0}$ constructed as explained in the Supplementary Note 3. Considering that all operators are diagonal in $M$, only the $\langle JM|\rho(t)|J'M\rangle$ terms were computed. From knowledge of these density matrix elements, the alignment factor was computed from

$$\langle \cos^2\theta - 1/3 \rangle = \sum_{J,M} \langle JM|\rho(t)\cos^2\theta|JM\rangle = \sum_{J,M,J'} \langle JM|\rho(t)|J'M\rangle\langle J'M|\cos^2\theta|JM\rangle$$

with $\langle J'M|\cos^2\theta|JM\rangle = \frac{2}{3}\sqrt{(2J+1)(2J'+1)}\begin{pmatrix}J & 2 & J' \\ 0 & 0 & 0\end{pmatrix}\begin{pmatrix}J & 2 & J' \\ M & 0 & -M\end{pmatrix}$, where (:::) is a 3J symbol. Note that this implies $J'$-$J$=0,$\pm$2.

**Data analysis**

For each delay $\tau_{12}$ between the laser pulses $P_1$ and $P_2$, the measured and computed time-dependent alignment factors for various gas densities $d$ were analyzed as follows. The variation with $d$ of the peak-to-dip amplitude $S$ of the echo (see Fig. 2b) was (nicely, see Fig. 2c) least-squares fitted by $S(d,\tau_{12}) = A(\tau_{12})\exp[-d/d_0(\tau_{12})]$ floating $A(\tau_{12})$ and $d_0(\tau_{12})$. A density-normalized decay time constant was then defined as $\tau_E(\tau_{12}) = 2\tau_{12}d_0(\tau_{12})$, considering the fact that the echo appears at $t$=2$\tau_{12}$ after the time origin defined by $P_1$. For the revivals, the same approach was used except that $P_2$ was turned off, and a density-normalized time constant was determined for each of them from the decay of their amplitude with density using $\tau_R(t_R) = t_R d_0(t_R)$, where $t_R$ is the time delay between the revival and $P_1$.

**Acknowledgments**
The work was supported by the ERDF Operational Programme – Burgundy 2014/2020 and the EIPHI Graduate School (contract "ANR-17-EURE-0002"). J. Ma acknowledges the support from the China Scholarship Council (CSC). J.-M. Hartmann benefited, for the computer simulations, from the IPSL mesocenter ESPRI facility which is supported by CNRS, UPMC, Labex L-IPSL, CNES, and Ecole Polytechnique. J.Wu acknowledges the support by the National Key R&D Program (Grant No. 2018YFA0306303) and the NSFC (Grant Nos. 11425416, 11761141004, and 11834004).

**Author contributions**
O.F. with J.-M.H. planned the project and wrote the manuscript. H.Z. and B.L., with the assistance from J.M. and F.B., designed and installed the experiment. J.M. carried out the measurements. B.L. and J.M. analyzed the data. J.-M.H. and C.B. developed the theory and carried out the simulations. All authors took part in regular discussions and contributed to the final manuscript.

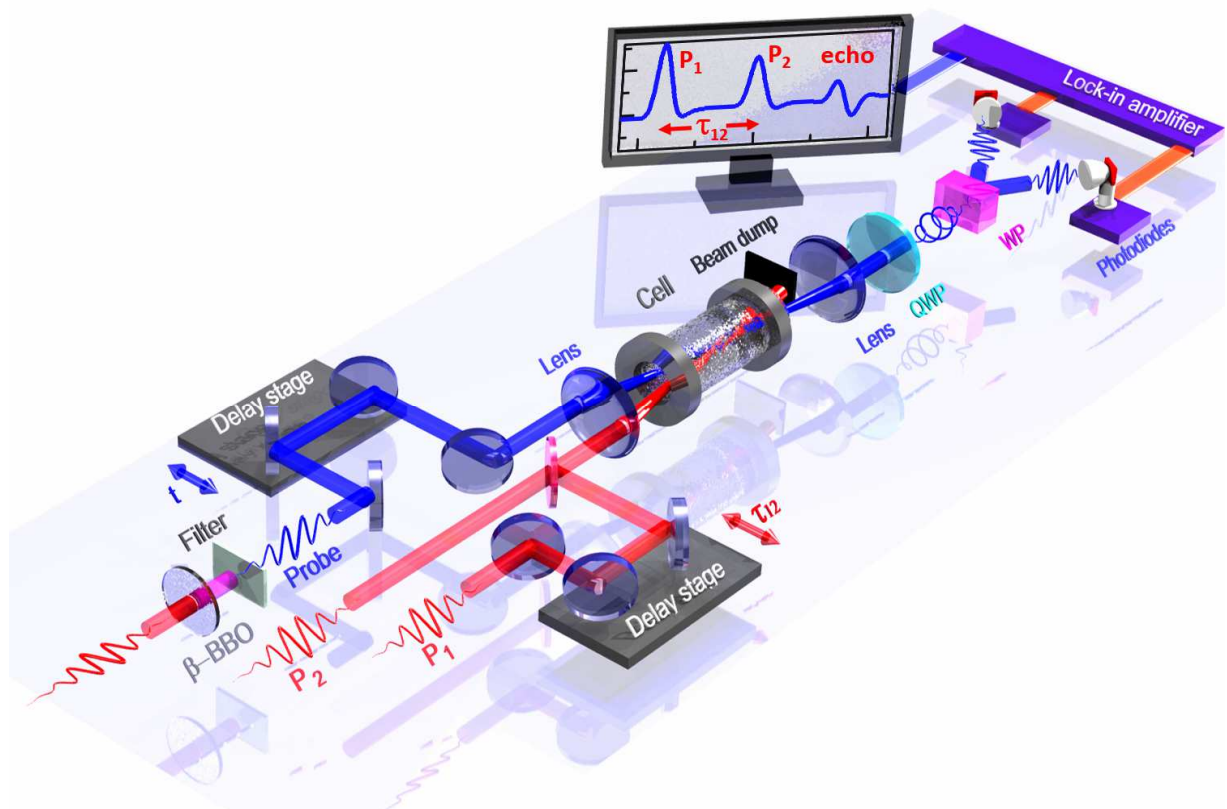

**Figure 1: Schematic representation of the experimental set-up.** N$_2$O gas molecules contained in a room-temperature high-pressure gas cell are impulsively aligned by two time-delayed 800 nm pulses P$_1$ and P$_2$. The induced rotational dynamics is measured through the time-dependent birefringence experienced by a 400 nm probe pulse. The detection uses two photodiodes connected head-to-tail to a lock-in amplifier delivering a signal proportional to the part of the probe field that has been depolarized by the aligned molecules. The image reproduced on the computer screen illustrates the alignment echo signal produced by the two time-delayed strong laser kicks; QWP, quarter-wave plate; WP, Wollaston prism.

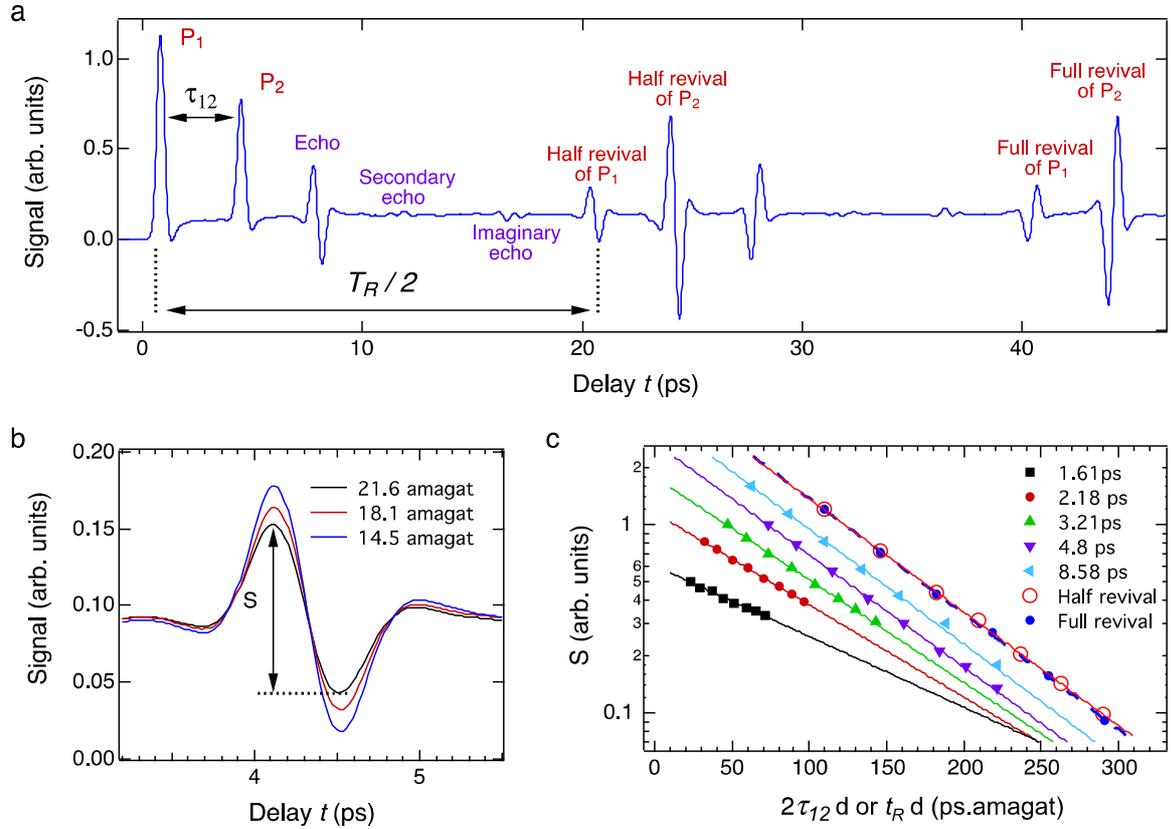

**Figure 2: Rotational structures of N$_2$O aligned by two laser kicks.** (**a**) Alignment signal recorded in pure N$_2$O gas at low pressure by scanning the temporal delay between the first aligning pulse P$_1$ (at $t$=0) and the probe pulse over slightly more than the full rotational period $T_R$=40.4 ps of the molecule. The peaks identified by P$_1$ and P$_2$ correspond to the transient alignment signals produced by the two pulses separated by the delay $\tau_{12}$. The main echo is generated at $t$=2$\tau_{12}$, with the secondary echo observable at $t$=3$\tau_{12}$, and the imaginary echo produced at $T_R$/2-$\tau_{12}$ (equivalent features also appearing at times shifted by +$T_R$/2). In addition to echoes, other transients corresponding to the standard half and full alignment revivals of P$_1$ (at $T_R$/2 and $T_R$) and P$_2$ (at $T_R$/2+$\tau_{12}$ and $T_R$+$\tau_{12}$) are also observed. (**b**) Alignment traces of N$_2$O diluted in He measured at various densities around the main echo at 2$\tau_{12}$ for $\tau_{12}$=2.18 ps. The amplitudes $S$ of the alignment structures are measured from peak to dip. (**c**) Amplitudes of the half revival (open red circles), full revival (full blue circles), and of the main echo for five different values of the delay $\tau_{12}$ versus the gas density $d$ multiplied by the time of observation $t_R$ ($T_R$/2 or $T_R$) and 2$\tau_{12}$ for the revivals and echo, respectively, expressed in picosecond amagat (ps.amagat, with 1 amagat=2.687 10$^{25}$ molec.m$^{-3}$) units. The lines indicate the best exponential fits.

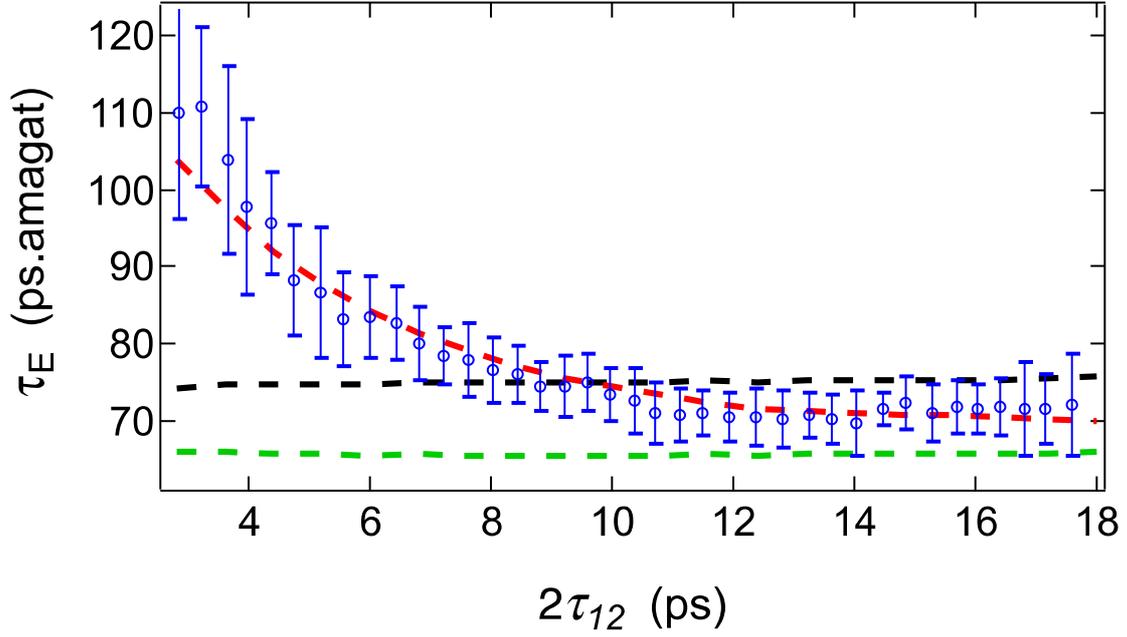

**Figure 3: Time constants of collisional dissipation of $N_2O$.** The blue circles with error bars (representing two standard deviations of the mean) are the density-normalized decay time constants $\tau_E$ of the echoes deduced from the measurements of the alignment signal recorded at various $N_2O(4\%)$+He(96%) gas densities and fixed delays $\tau_{12}$ between the two pulses. The dashed lines denote the results of simulations conducted by solving the density matrix equations for molecules impulsively aligned by two short laser pulses and interacting with each other through collisions. The green and black dashed lines have been obtained using the standard Bloch equations (i.e. using the secular approximation) with initial $N_2O$ rotational populations corresponding to temperatures of $T$=100 K and 295 K, respectively. The red dashed line represents the results obtained, for populations associated with $T$=100 K (see text), using the nonsecular Redfield equations, i.e. including all relaxation terms in equation (3).

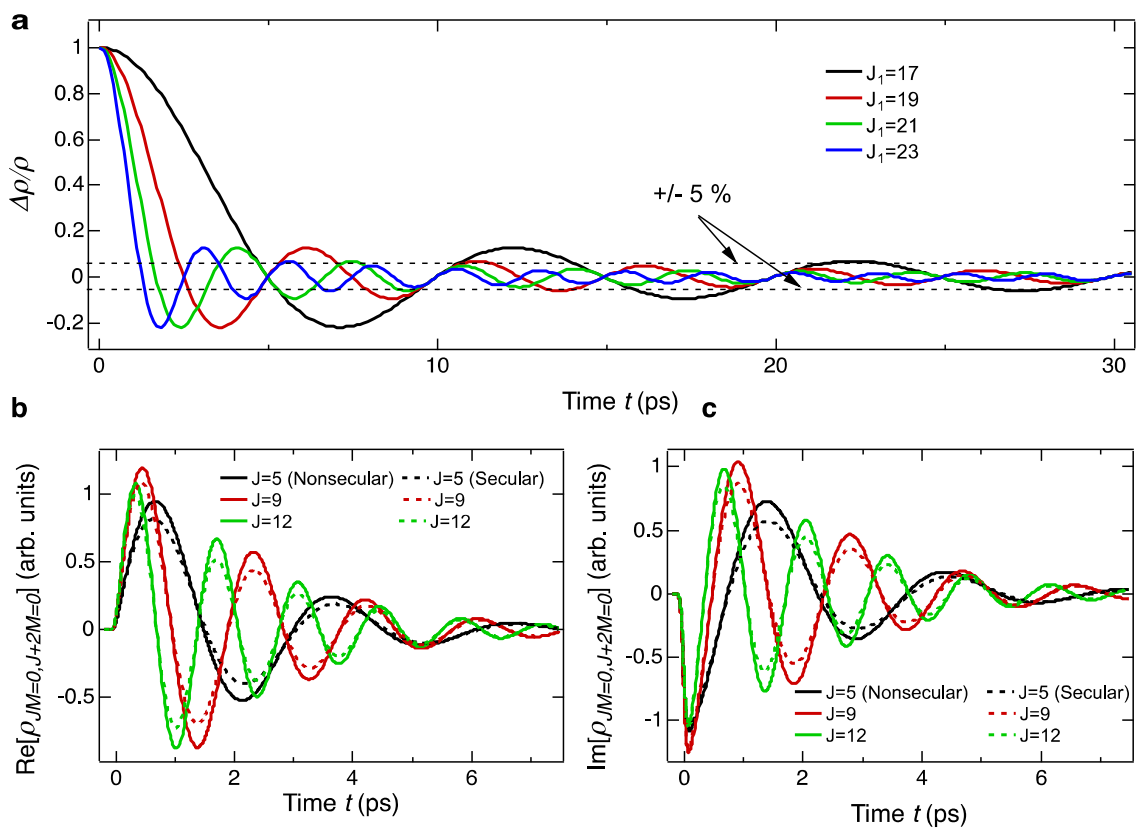

**Figure 4: Secular and nonsecular effects.** (**a**) Relative modification of the $\rho_{J_0M_0,J_0+2M_0}(t)$ coherence for $J_0$=15 induced by collisional transfers from the coherences $\rho_{J_1M_1,J_1+2M_1}(t)$ with $J_1$=17 (black), $J_1$=19 (red), $J_1$=21 (green), and $J_1$=23 (blue), all normalized to unity at $t$=0. This simple modeling (see Supplementary Note 2) of coherence transfers considering only few rotational states around the most populated state of $N_2O$ at 300 K reveals that during the short time evolution of the system, exchanges between coherences are efficient, which slow down the decay of the alignment factor with respect to what would be obtained using the secular approximation that neglects these collisional transfers and only takes the losses into account. (**b**) and (**c**) Computed real and imaginary parts of the $\rho_{JM=0,J+2M=0}(t)$ coherences for $N_2O$ in He at 33 amagat, obtained using the nonsecular (full lines) and secular (dashed lines) models.

# Supplementary Information for

"Observing collisions beyond the secular approximation limit"

Ma et al.

**Supplementary Note 1: On the Markovian approximation**
The Markovian approximation for rotational relaxation and decoherence processes assumes that the quantity of interest (here the time dependence, through the time evolution of the density matrix, of the alignment factor) can be modeled by considering that all collisions are complete. Also called impact approximation, it thus neglects the memory effects associated with those collisions that are on-going at the time ($t=0$) of the excitation of the system or at that ($t=2\tau_{12}$, for the echo) where the system is observed. This implicitly assumes that the relative number of on-going collisions is very small and/or that they have an effect over a very short interval of time. For the system of $N_2O$ gas highly diluted in He investigated in the present study, both criteria are well satisfied so that the Markovian approximation is valid. Indeed, the $N_2O$-He accurate ab initio intermolecular potential [1] has an almost negligible well (with respect to the mean kinetic energy at room temperature). The repulsive front is relevant (i.e. has values between 0 and 600 K) in the [$R_0$-$\Delta R$, $R_0$] range of intermolecular distances where $R_0$ varies between 2.7 and 4.2 A depending on the respective orientations of the molecular and intermolecular axes and $\Delta R$ is typically 0.5 Å. This means that, for a $N_2O$ molecule to be colliding with a He atom, the latter should be in the interval between the two spheres centered on the $N_2O$ center of mass and of typical radii 3.0 and 3.5 Å. This corresponds to a very small volume (about 70 Å$^3$) that statistically contains very few He atoms unless the gas density is very high [for 15 amagat which corresponds to 4 $10^{26}$ atom.m$^{-3}$, this statistical number is about 0.025). In addition, the mean relative speed of $N_2O$-He collisions at 295 K is 1300 m.s$^{-1}$, which implies that the distance between the two spheres (0.5 Å) is, on average, traveled in 0.038 ps. The above given numbers show that non-Markovian effects are of very small amplitude and only exist at time scales much shorter than all those involved in the present study. Assuming Markovian collisions is thus a very good approximation for the $N_2O$-He system but this would not be the case for pairs of colliders involving much smaller relative speeds and/or intermolecular forces at much longer ranges.

**Supplementary Note 2: On the validity of the secular approximation**
Consider a case where coherences have been instantaneously generated by laser pulse at $t=0$. This creates non-zero off-diagonal elements $\rho_{JM,J'M}(t=0^+)$, which, in the absence of collisions, have the expression:

$$\rho_{JM,J'M}(t) = \rho_{JM,J'M}(t=0^+)\exp[-i(E_J - E_{J'})t/\hbar]. \quad (1)$$

Let us now assume that collisions induce transfers between coherences and, in particular, from $\rho_{J_1M_1,J_1'M_1}(t)$ to $\rho_{J_0M_0,J_0'M_0}(t)$ with a real-valued density-normalized rate $K_{J_1M_1,J_1'M_1 \to J_0M_0,J_0'M_0}$. For a gas density $d$, the total change of $\rho_{J_0M_0,J_0'M_0}(t)$ due to transfers from $\rho_{J_1M_1,J_1'M_1}(t)$ at all possible times $t_0 > 0$ is:

$$\Delta\rho_{J_0M_0,J_0'M_0}(t) = K_{J_1M_1,J_1'M_1 \to J_0M_0,J_0'M_0} d \times \int_0^t \rho_{J_1M_1,J_1'M_1}(t_0) e^{-i(E_{J_0}-E_{J_0'})(t-t_0)/\hbar} dt_0 \; . \quad (2)$$

If we assume that these transfers are relatively small, we can consider that $\rho_{J_1M_1,J_1'M_1}(t_0)$ remains unchanged when $t_0$ varies so that Supplementary Equation (1) can be used and Supplementary Equation (2) becomes:

$$\Delta\rho_{J_0M_0,J_0'M_0}(t) = \rho_{J_1M_1,J_1'M_1}(t=0^+) K_{J_1M_1,J_1'M_1 \to J_0M_0,J_0'M_0} d$$
$$\times \exp\left[-i(E_{J_0} - E_{J_0'})t/\hbar\right] \int_0^t e^{-i(E_{J_1} - E_{J_1'} - E_{J_0} + E_{J_0'})t_0/\hbar} dt_0 \quad , (3)$$

i.e., using Supplementary Equation (1) again

$$\frac{\Delta\rho_{J_0M_0,J_0'M_0}(t)}{\rho_{J_0M_0,J_0'M_0}(t)} = A \times d \frac{1 - \exp\left[-i(E_{J_1} - E_{J_1'} - E_{J_0} + E_{J_0'})t/\hbar\right]}{i(E_{J_1} - E_{J_1'} - E_{J_0} + E_{J_0'})/\hbar} \quad , (4)$$

with

$$A \equiv \frac{\rho_{J_1M_1,J_1'M_1}(t=0^+)}{\rho_{J_0M_0,J_0'M_0}(t=0^+)} K_{J_1M_1,J_1'M_1 \to J_0M_0,J_0'M_0} \quad . (5)$$

For a study of the relative evolution of this quantity with time, one must normalize the effect of collisions, which implies that:

$$d \times t = \text{Constant} \equiv \delta \quad , (6)$$

so that Supplementary Equation (4) becomes:

$$\text{Re}\left[\frac{\Delta\rho_{J_0M_0,J_0'M_0}(t)}{\rho_{J_0M_0,J_0'M_0}(t)}\right] = B\,\text{sinc}\left[(\omega_{J_1J_1'} - \omega_{J_0J_0'})t\right] \quad , (7)$$

where Re[...] denotes the real part, sinc[...] designates a cardinal sine, $\omega_{JJ'} \equiv (E_J - E_{J'})/\hbar$ and $B \equiv \delta.A$.

## Supplementary Note 3: Rates of population and coherence transfers

### 3.1 Infinite Order Sudden model

The elements $\langle J_1M_1, J_1'M_1 | \Lambda | J_0M_0, J_0'M_0 \rangle$ of the matrix describing losses and transfers among populations $[\rho_{JM,JM}(t)]$ and coherences $[\rho_{JM,J'M}(t)$ with $J \neq J']$ were computed using the Infinite Order Sudden Approximation (IOSA). Recall that the IOSA is obtained [2] from the Close-Coupling model by first making the Centrifugal Sudden (or Coupled States CS) approximation, i.e. by assuming an effective orbital momentum eigenvalue and freezing the centrifugal potential. Then, starting from the CS equations, the IOSA freezes the molecular rotation during each collision, which corresponds to neglecting the energy difference between different rotational states. For molecule-atom collisions, one can then separate the collisional cross-sections into spectroscopic and dynamical factors [3,4]. For the $N_2O$-He collisions considered here, this is valid thanks to the high value of the $N_2O$-He pair relative translational speed and the fact that the intermolecular forces for this system are significant only in a very narrow range of intermolecular distances [1] (see also Supplementary Note 1). Within this frame, the relaxation matrix elements describing collisional exchanges among and between populations and coherences are given by [5,6]:

$$\langle J_1M_1, J_1'M_1 | \Lambda | J_0M_0, J_0'M_0 \rangle = -\sqrt{(2J_1+1)(2J_1'+1)(2J_0+1)(2J_0'+1)}$$
$$\sum_L \begin{pmatrix} J_0 & J_1 & L \\ 0 & 0 & 0 \end{pmatrix} \begin{pmatrix} J_0' & J_1' & L \\ 0 & 0 & 0 \end{pmatrix} \begin{pmatrix} J_0 & J_1 & L \\ M_0 & -M_1 & M_1-M_0 \end{pmatrix} \begin{pmatrix} J_0' & J_1' & L \\ M_0 & -M_1 & M_1-M_0 \end{pmatrix} (2L+1)Q_L \quad , (8)$$

where (:::) is a 3J symbol and $Q_L$ is the rate of collisional de-excitation from level $J=L$ to level $J=0$. As widely done before for other types of studies [7,8], the latter are modelled using the exponential power gap law based on three parameter (A,α,β), i.e.:

$$Q_{L\neq 0} = A[L(L+1)]^{-\alpha} \exp(-\beta E_L / k_B T) \,, \quad (9)$$

where $k_B$ is the Boltzmann constant and, according to Ref. [9]:

$$Q_0 = - \sum_{L\neq 0} (2L+1) Q_L \,. \quad (10)$$

Note that this convention for the $L=0$ term differs from that used in our previous studies [7,10].

### 3.2 Input data

As widely done before [7,11], the parameters of Supplementary Equation (9) were here deduced from fits of pressure broadening coefficients of infrared lines. Indeed, for the $J_i \to J_f$ optical transition, this coefficient $\gamma_{J_i J_f}$ is, within the IOSA, given by [7,9,12]:

$$\gamma_{J_i J_f} = -Q_0 + (2J_i+1)(2J_f+1) \\ \times \sum_{L\neq 0} \begin{pmatrix} J_i & J_i & L \\ 0 & 0 & 0 \end{pmatrix} \begin{pmatrix} J_f & J_f & L \\ 0 & 0 & 0 \end{pmatrix} \begin{Bmatrix} J_i & J_f & 1 \\ J_f & J_i & L \end{Bmatrix} \times (2L+1) Q_L \,, \quad (11)$$

where {:::} is a 6J symbol. The least-square fit of the measured He-broadened widths of $N_2O$ lines of Ref. [13] using Supplementary Equations (9)-(11) leads to A=22.8 $10^{-3}$ cm$^{-1}$atm$^{-1}$ (4.66 $10^{-3}$ ps$^{-1}$amagat$^{-1}$), α=1.05, and β=0.04 with the good agreement shown in Supplementary Figure 1. Note that only the even values of $L$ are retained in Supplementary Equations (8), (9), and (11), as done in Ref. [11], since the $N_2O$-He intermolecular potential is almost insensitive to a rotation by π of the $N_2O$ molecule [1].

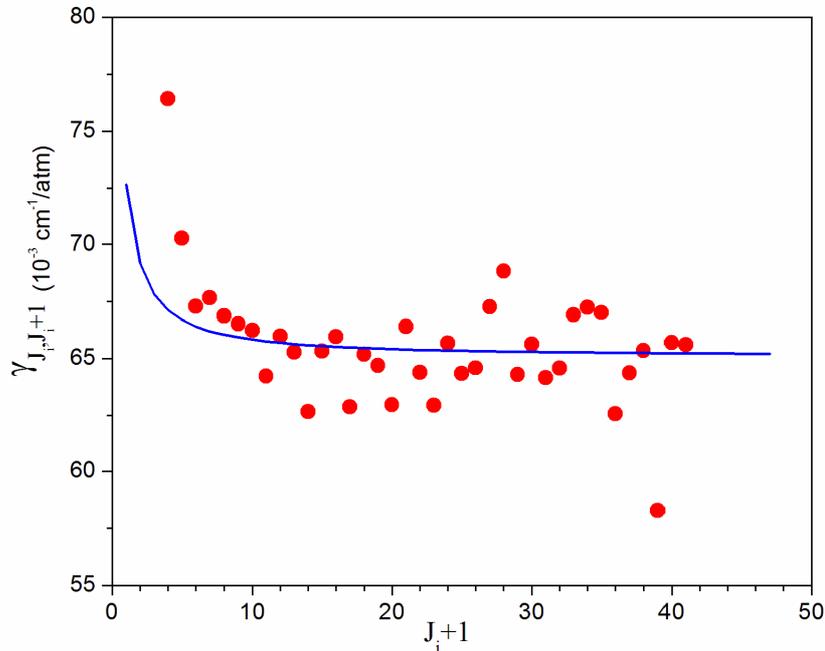

**Supplementary Figure 1: Pressure broadening of $N_2O$ lines by He**. Comparison between measured line broadening coefficients of $N_2O$ by He for infrared R($J_i \to J_i+1$) lines [13] (symbols) [converted from 303 K to 295 K by a factor $(303/295)^{1/2}$] with those fitted with Supplementary Equations (9)-(11) (line).

### 3.3 From nonsecular to secular

Within the secular approximation, only the terms $\langle JM, J'M | \Lambda | JM, J'M \rangle$ driving the loss of coherence ($J \neq J'$) and population ($J = J'$) and those $\langle J'M', J'M' | \Lambda | JM, JM \rangle$ associated with exchanges between the $|JM\rangle$ and $|J'M'\rangle$ populations are kept. Let us first consider the exchanges between populations, driven by $\langle J'M', J'M' | \Lambda | JM, JM \rangle$ with $|JM\rangle \neq |J'M'\rangle$. The associated rate is straightforwardly obtained from Supplementary Equation (8):

$$K_{J,M \to J',M'} = -\langle J'M', J'M' | \Lambda | JM, JM \rangle$$
$$= (2J+1)(2J'+1) \sum_{L \neq 0} \begin{pmatrix} J & J' & L \\ M & -M' & M'-M \end{pmatrix}^2 \begin{pmatrix} J & J' & L \\ 0 & 0 & 0 \end{pmatrix}^2 (2L+1) K(L \to 0) \quad , (12)$$

which is identical to Equation (A2) of Ref. [10]. Now concerning the rate $\langle JM, JM | \Lambda | JM, JM \rangle$ driving the loss of the $|JM\rangle$ population, it can be transformed using 3J properties and Supplementary Equation (10), leading to

$$\langle JM, JM | \Lambda | JM, JM \rangle = \sum_{(J',M') \neq (J,M)} K_{J,M \to J',M'} \quad , (13)$$

in agreement with Equation (19) of Ref. [10]. Finally, the loss of the $|JM, J'M\rangle$ coherence is driven by the term $\langle JM, J'M | \Lambda | JM, J'M \rangle$. Again using properties of the 3J symbols and Supplementary Equation (10), it can be rewritten as:

$$\langle JM, J'M | \Lambda | JM, J'M \rangle = \frac{1}{2} \left[ \sum_{(J'',M'') \neq (J,M)} K_{J,M \to J'',M''} + \sum_{(J'',M'') \neq (J',M)} K_{J',M \to J'',M''} \right]$$
$$+ \sum_{L \neq 0} \frac{1}{2} \left\{ (2J+1) \begin{pmatrix} J & L & J \\ 0 & 0 & 0 \end{pmatrix} \begin{pmatrix} J & L & J \\ -M & 0 & M \end{pmatrix} - (2J'+1) \begin{pmatrix} J' & L & J' \\ 0 & 0 & 0 \end{pmatrix} \begin{pmatrix} J' & L & J' \\ -M & 0 & M \end{pmatrix} \right\}^2 (2L+1) Q_L \quad , (14)$$

which is identical to Equations (12) and (16) of Ref. [10]. Finally note that assuming that $K_{J,M \to J',M'}$ is independent of $M$ and $M'$ leads [10] to the equations used in Refs. [14,15].

### Supplemenraty References